# Characterization of the Frictional Response of Squamata Shed Skin in Comparison to Human skin

H. A. Abdel-Aal [1]  M. El Mansori
Arts et Métier ParisTech, Rue Saint Dominique BP 508,
51006 Chalons-en-Champagne,
France
[1] *corresponding author:* <u>Hisham.abdel-aal@chalons.ensam.fr</u>

**ABSTRACT**

Deterministic surfaces are constructs of which profile, topography and textures are integral to the function of the system they enclose. They are designed to yield a predetermined rubbing response. Developing such entities relies on controlling the structure of the rubbing interface so that, not only the surface is of optimized topography, but also is able to self-adjust its behavior according to the evolution of sliding conditions. Inspirations for such designs are frequently encountered in natural species. In particular, and from a tribological point of view, Squamate Reptiles, offer diverse examples where surface texturing, submicron and nano-scale features, achieves frictional regulation. In this paper, we study the frictional response of shed skin obtained from a Python regius snake. The study employed a specially designed tribo-acoustic probe capable of measuring the coefficient of friction and detecting the acoustical behavior of the skin in vivo. The results confirm the anisotropy of the frictional response of snakes. It is found that the coefficient of friction depends on the direction of sliding: the value in forward motion is lower than that in the backward direction. Diagonal and side winding motion induces a different value of the friction coefficient. We discuss the origin of such a phenomenon in relation to surface texturing and study the energy constraints, implied by anisotropic friction, on the motion of the reptile and to establish a reference for comprehending the frictional response we draw a comparison to the friction of human skin.

## 1. Introduction

The design of ultra precise, structured surfaces for improved lubrication is an active area of research. In seeking inspirations for such custom designs many engineers turn toward natural systems because of their advantageous features which include superior functionality, optimal energetics, and harmony between shape form and function. From a tribology perspective, the existence of deterministic surfaces, i.e., surfaces that are an integral design feature, of the species, that facilitate function and optimize performance is a point of deserving interest.

Generation of design in natural systems (geometry, pattern, form and texture) is a holistic phenomenon that synchronizes all design constituents toward an overall optimized performance envelope. Such an approach yields deterministic design outputs that while conceptually simple, are targeted toward minimum energy expenditure. Natural engineering, thus, seeks trans-disciplinary viable alternatives that, given functional constraints, require minimum effort to construct and economizes effort while functioning. An analogous design paradigm, within the Man Engineered Systems, *MES*, domain, has not matured as of yet. This is primarily due to a dominant design culture that attempts to tame nature rather than to establish harmonious co-existence. Clearly, generation of deterministic surface designs of predictable tribo-performance profile is a challenge within such model. As such, probing the essence of surface design generation in natural systems, especially within bio-species, is of great benefit. This is because biological materials, through million years of existence, have evolved optimized topological features that enhance wear and friction resistance [1]. One species of remarkable tribological performance that may serve as an inspiration for deterministic surface texturing is that of snakes.
.
Snakes lack legs and use the surface of the body itself to generate propulsion on the ground during locomotion. For such a purpose, frictional tractions are necessary, in order to transmit forces to the ground. Depending on the snake species, type of movement, environment and preferred substrate, different parts of the body must have different functional requirements and therefore different frictional properties. That is the snake species, is a true representative of a heterogeneous tribo-system with a high degree of functional complexity, despite which, they don't suffer damaging levels of wear and tear.

Many researchers investigated the intriguing features of the serpentine family. Adam and Grace **[2]** studied the ultra structure of pit organ epidermis in Boid snakes to understand infrared sensing mechanisms. Johnna et al **[3]** investigated the permeability of shed skin of pythons (python molurus, Elaphe obsolete) to determine the suitability as a human skin analogue. Mechanical behavior of snake skin was also a subject of several studies as well. Jayne **[4]** examined the loading curves of six different species in uni-axial extension. His measurements revealed substantial variation in loads and maximum stiffness among samples from different dorsoventral regions within an individual and among homologous samples from different species. Rivera et al.**[5]** measured the mechanical properties of the integument of the common garter snake (Thomnophis sirtalis-Serpentine Colubridae). They examined mechanical properties of the skin along the body axis. Data collected revealed significant differences in mechanical properties among regions of the body. In particular, and consistent with the demands of macrophagy, it was found that the pre-pyloric skin is more compliant than post pyloric skin. Prompted by needs to design bio-inspired robots several researchers probed the frictional features of snake motion to understand the mechanisms responsible for regulating legless locomotion. Hazel et al **[6]** used AFM scanning to probe the nano-scale design features of three snake species. The studies of Hazel and Grace revealed the asymmetric features of the skin ornamentation to which both authors attributed frictional anisotropy.

In order to mimic the beneficial performance features of the skin, an engineer should be provided with parametric guidelines to aid with the objective-oriented design process. These should not only be dimensional. Rather, they should extend to include metrological parameters used to characterize tribological performance of surfaces within the MES domain. Thus, in order to deduce design rules there exists a need for quantification of the relationship governing micro-structure and strength topology of the bio-surface; exploring the quantitative regulation of macro and micro texture, and finally devising working formulae that describe (and potentially predict) load carrying capacity during locomotion in relation to geometrical configuration at both the micro and the macro scale. This paper is a preliminary step toward that goal.

In this work, we apply a multi-scale surface characterization approach to probe the geometrical features of shed skin obtained from a Ball Python (Python Regius). Such a species is of tribological interest because of its locomotion taking place within a non-breakable boundary lubrication regime. This feature is facilitated through the topology of skin ornamentation which renders the species of interest to industrial surface texture applications, such as honing where surfaces are designed for minimal lubricant consumption and for designers of hip and knee prostheses where maintaining a continuous boundary lubrication regime is a must. The emphasis in the current work, therefore, is on deducing those metrological aspects, and tribological features of the shed skin that are deemed essential to quantify the quality of tribo-performance.

The paper is organized as follows: In the first part we provide background information about the species under study, its biological features, skin morphology, and essential features of the skin shedding process. This is followed in the second section of the manuscript by reporting optical and Scan Electron Microscopy (SEM) observations of the shed skin. Further, we report within the second section, on the metrological aspects of shed skin surface topography along with a comparison to the topology of human skin. Finally, in the third section of the paper, we provide a comparison between friction coefficient (COF) measurements of the shed skin and values of the COF for human skin reported in open literature.

## 2. Background
### 2.1. The Python species

*Python Regius* is a non-venomous python species native to Africa. It is the smallest of the African pythons and is popular in the pet trade. The name *ball python* refers to the animal's tendency to curl into a ball when stressed or frightened **[7]**. Adults generally do not grow to more than 90-120 cm long. Females tend to be slightly bigger than males maturing at an average of 120-150 *cm*. Males usually average around 90-105 cm. The build is stocky while the head is relatively small. The color pattern is typically black with light brown-green side and dorsal blotches. The belly is white or cream that may or may not include scattered black markings.

### 2.2. general features

Snakes, like other reptiles, have a skin covered in scales. They are entirely covered with scales or scutes of various shapes and sizes. Scales protect the body of the snake, aid it in locomotion, allow moisture to be retained within, alter the surface characteristics such as roughness to aid in camouflage, and in some cases even aid in prey capture. The simple or complex colorations patterns (which help in camouflage and anti-predator display) are a property of the underlying skin, but the folded nature of scaled skin allows bright skin to be concealed between scales then revealed in order to startle predators. Snake scales are formed by the differentiation of the snake's underlying skin or epidermis. Snakes periodically molt their scaly skins and acquire new ones. Each scale has an outer surface and an inner surface. A snake hatches with a fixed number of scales. The scales do not increase in number as the snake matures nor do they reduce in number over time. The scales however grow larger and may change shape with each molt.

### 2.3. Structure of snake skin

Reptile skin, like that of many vertebrates, has two principal layers: the *dermis* which is the deeper layer of connective tissue with a rich supply of blood vessels and nerves; and the *epidermis,* which in reptiles consists of up to seven sub-layers or "*strata*" of closely packed cells, forming the outer protective coating of the body **[8]**. The "epidermis" has no blood supply, but its' inner most living cells obtain their nourishment by the diffusion of substances to and from the capillaries at the surface of the "dermis" directly beneath them. There are seven epidermal layers, figure (1): *1*-the *"stratum germinativum"*, the deepest layer lining cells which have the capacity for rapid cell division and the six layers which form each *"epidermal generation"*, the old and the new skin layers. These are: *2-3*: the clear layer and the lacunar layer, which matures in the old skin layer as the new skin is growing beneath.

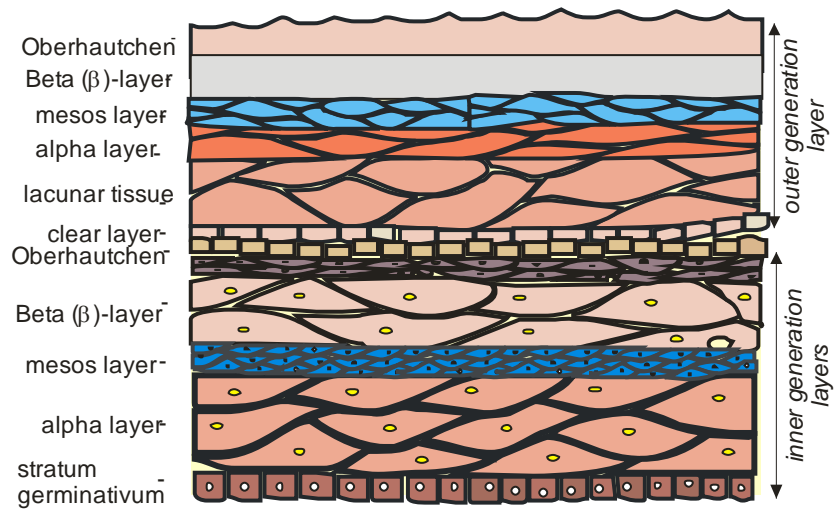

Figure 1 Generalized epidermis of a squamate reptile.

*4-6*: the alpha ($\alpha$)–layer, the mesos layer and the beta ($\beta$)-layer, these layers consist of cells which are becoming keratinized with the production of two types of keratin ($\alpha$ and $\beta$ keratin)d . These cells are thus being transformed into a hard protective layer. *7*- The *"oberhautchen"* layer, which forms the toughest outer most layer of keratinized dead skin cells.

## 2.4. Skin shedding

In most mammals, the structure of the epidermis is less complex and the outermost dead skin cells are constantly flaking off; this protective layer is constantly replaced from below. The deepest layer of cells, the *"stratum germinativum "* , is constantly dividing and multiplying, and so the layers are on the move outwards **[9]**. In reptiles, however, this cell division, in the *"stratum germinativum "* , only occurs periodically **[10]**, and when it does all the layers above it in the area where the cell division occurs are replaced entirely. That is the, the reptile, grows a second skin underneath the old skin, and the "sheds" the old one. About two weeks before the reptile sheds its skin, the cells in the *stratum germinativum* begin active growth and a second set of layers form slowly underneath the old ones. At the end of this time the reptile effectively has double skin. Following such a process, the cells in the lowest layers of the old skin, the clear and the lacunar layers, and the Oberhautchen layer of the skin below undergo a final maturation and a so called *"shedding complex forms"*. Fluid is exuded and forms a thin liquid layer between them. This gap between the two skins gives a milky appearance to a shedding reptile. Enzymes, in this fluid, break down the connections between the two layers. The old skin lifts and the reptile actively removes it.

## 3. Observation of shed skin

All observations reported herein pertain to shed skin obtained from a 115 cm, 14 years old male Ball Python (Python Regius) housed individually in a glass container with news paper substrate. For optical microscopy observations skin was observed as is, whereas all samples for Scan Electron Microscopy observations were coated by a vacuum deposited Platinum layer of thickness 10 nm. Surface topography analysis took place through two methods: SEM imaging in topography mode and through examination using a white light interferometer (WYKO NT3300 3D Automated Optical Profiling System).

### 3.1. Optical Microscopy Observations

Initial observations, using optical microscopy, of scale structure took place without any treatment of the skin. Figure (2) depicts the structure of the scales at two positions within the skin in a region close to the waist of the snake. The first was from the back (dorsal scale) whereas the second position represented the belly of the snake (ventral). Note that although the general form of the cells is quite similar for both positions the size of a unit cell within the skin is quite different in both cases. In particular the cell is wider for the ventral (belly) positions. Each cell (scale) is also composed of a boundary and a membrane like structure. Note also the overlapping geometry of the skin and the scales (the so called scale and hinge structure). The skin from the inner surface hinges back and forms a free area that overlaps the base of the next scale which emerges below this scale figure (3).

### 3.2. Scan Electron Microscopy (SEM) Observations

Four positions on the shed skin were identified for initial examination. The choice of the positions was based on the functional profile of each position in the live species (figure 4). Position I is a representative of the neck region, position II represents the beginning of the trunk (waist) region, position III marks the boundary between the trunk and the tail region, finally position IV represents the tail region (containing the so called *subcaudal scales*). Skin swatches from each of the chosen positions were examined at different magnifications (X250-X15000) in topography mode. The surface of each sample was metalized by depositing a 10nm thin layer of platinum (Pt) using a sputter coater EMITECH K575X.

Figure 2  The structure of the scales on the inside of the shed skin at a region close to the mid section of the species at two orientations: back (dorsal) and abdominal (ventral).

Figure 3.  The details of dorsal scales from the inside of shed skin.  The terminology used is: membrane to denote the major area of the scale and boundary to denote the raised part forming the circumference of the scale. Scale marker is 1mm.

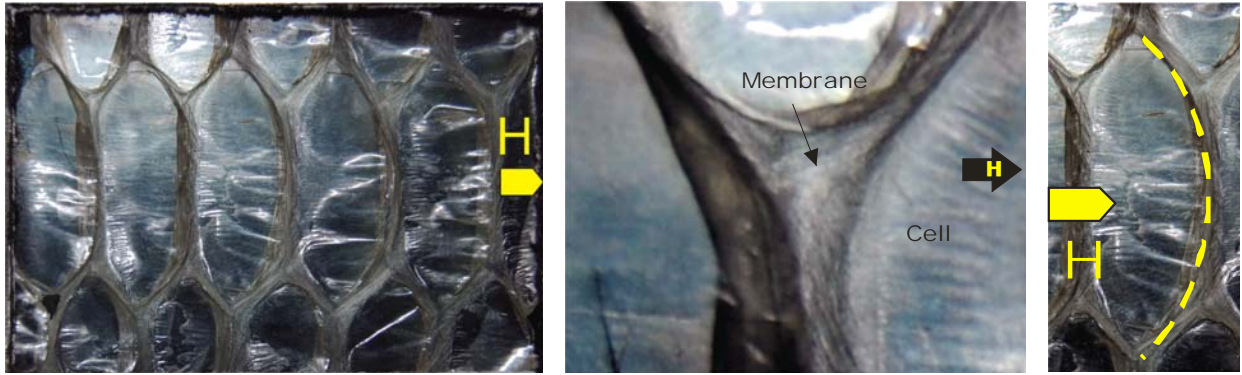

Figure 4    Equivalent positions chosen on the snake shed skin for SEM observations

For each position, samples from the dark and the light colored skin (see figure 6) were also examined along with samples from the underside of the body. Major features of the observations are shown in figures 5 and 6.

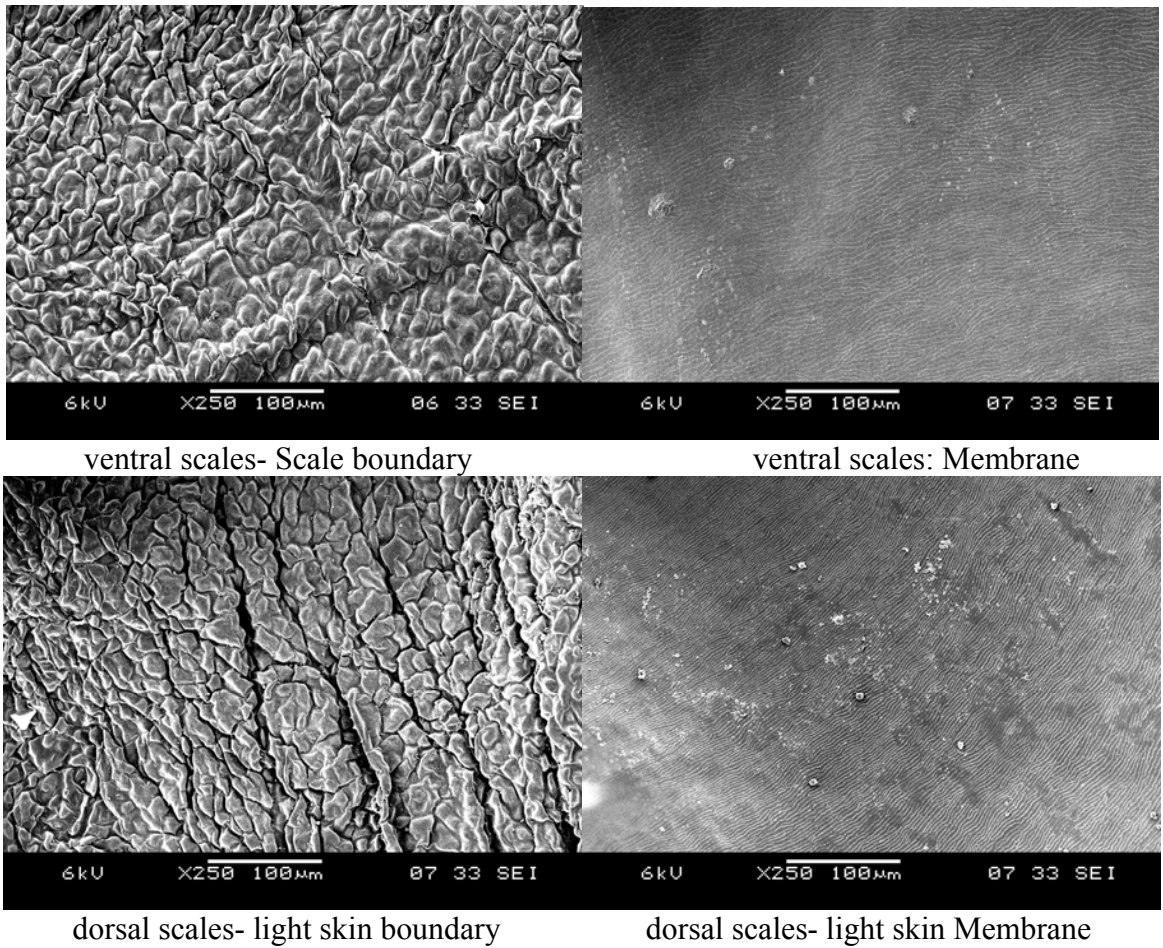

| ventral scales- Scale boundary | ventral scales: Membrane |
| dorsal scales- light skin boundary | dorsal scales- light skin Membrane |

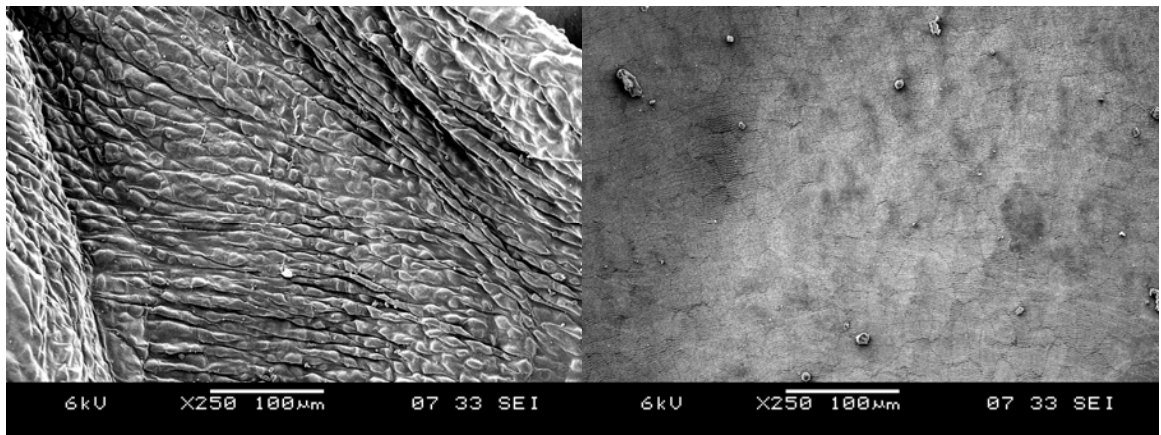

    dorsal scales- dark skin boundary      dorsal scales- dark skin Membrane

Figure 5 Major features of SEM observations of the skin swatches (X-250, scale marker 100 μm). Pictures orientation: head (top) tail (bottom).

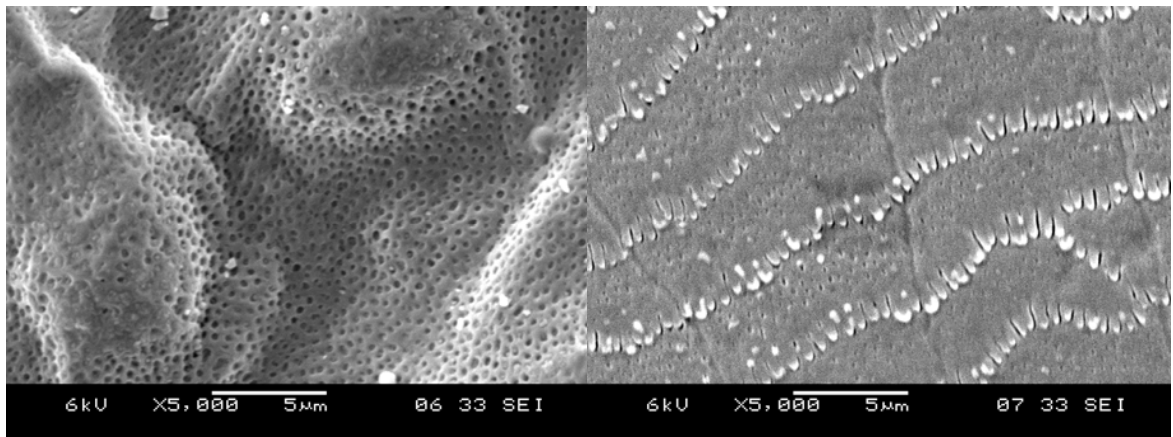

    ventral scales- Scale boundary      ventral scales: Membrane

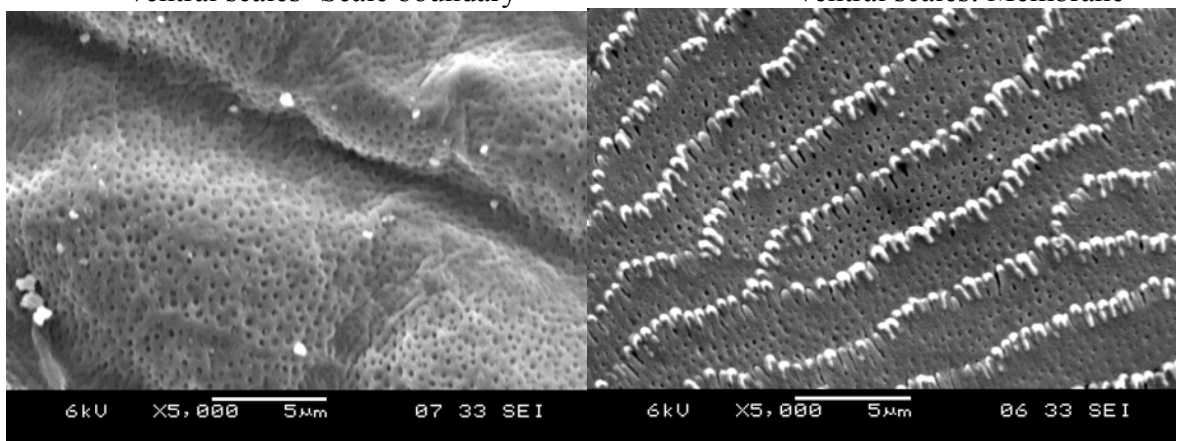

    dorsal scales- light skin boundary      dorsal scales- light skin Membrane

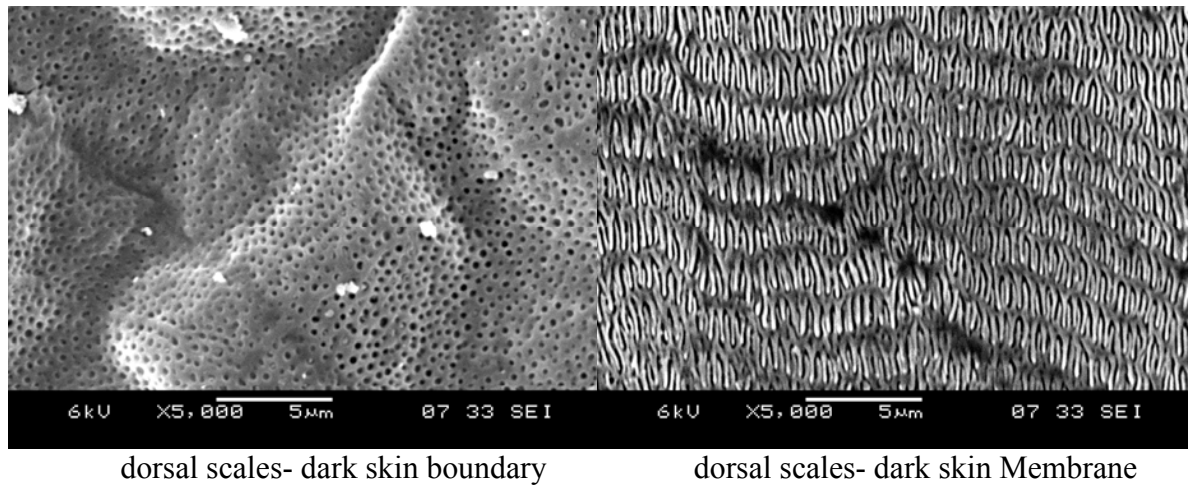

      dorsal scales- dark skin boundary        dorsal scales- dark skin Membrane

Figure 6   Major features of SEM observations of the skin swatches.(X-5000, scale marker 5 μm).  Picture orientation: head (top) and tail (bottom).

Note the inner structure that comprises pores. Two types of pores (or micro pits) may be distinguished: those located within the boundary and those located within the membrane. Image analysis of the pictures indicates that the diameter of the boundary-pores ranges between (200 nm – 250 nm).  The diameter of the membrane-pores was estimated by Hazel et al **[6]** using AFM analysis to be in the range (50nm -75 nm).

     Surface protrusions are also noted within the boundary.  These protrusions are of an asymmetric shape and irregular distribution.  The surface of the membrane also comprises micro-nano fibrile structures.  These are not of consistent shape and spacing.  Note for example that the shape of fibril located in the dark colored skin region is different than that located within the light colored skin region (compare the X-5000 pictures).  Moreover the density of the fibrils seems to be different within the different color regions (denser within the dark colored region).

     Figure (7) depicts the variation in the spacing between fibril rows $\lambda$. The figure presents measurements along twelve positions within the body of the snake.  The locations selected for measurements are represented in non-dimensional scale with respect to the total length of the snake (approximately 150 cm).   Each point, in the plot, represents an average of ten measurements of the parameter $\lambda$ taken from SEM images, X-3000, of the different skin colors and scale positions (i.e., dorsal light and dark skin in addition to dorsal scales). The figure indicates that the spacing between rows of fibrils differs by skin color and position within the body (ventral> dorsal light skin> dorsal dark skin). Additionally, the scatter in the value of $\lambda$ for the ventral scales is more pronounced than that for the dorsal scales (both light and dark skin). For ventral scales, the intra-spacing between fibrils in the head and the tail sections are the smallest, whereas spacing within the same positions on the dorsal scales are the highest.

     Further analysis of images revealed that the density of the boundary-pores vary by position. That is the number of pores per unit area is not constant along the body, rather it changes relative to the position. Figure (8) is a plot of the variation in the density of the pores

relative to the two sides of the skin (back-Dorsal scales and abdominal-Ventral scales) and in relation to the color of the skin (Light Patches (L) Vs Dark Patches D) within the back also.

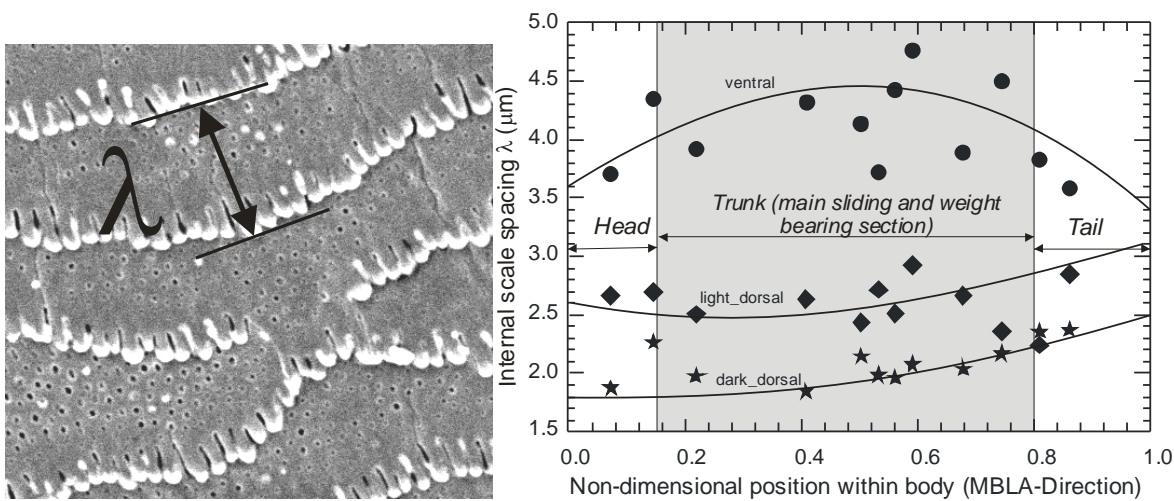

Figure 7: variation in the intra-spacing between rows of the micro fibrils located within the membrane (hinge) of the snake scales as a function of: distance along the MBLA, position of scales and color.

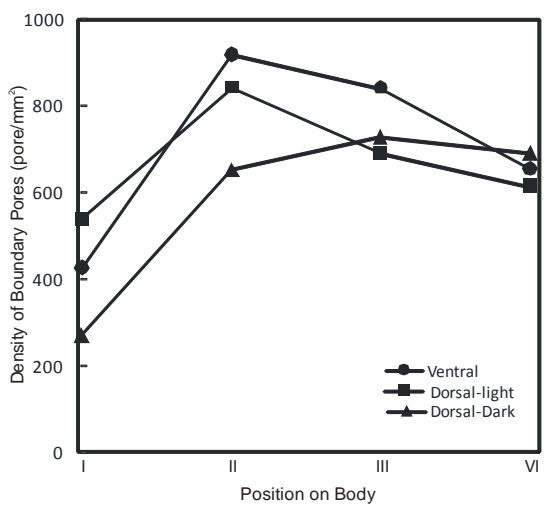

Figure 8  The variation in the density of the boundary pores (pore /mm$^2$) with position, and with color of skin.

### 3.3.   Metrology of the Surface

Examination of the surface topography features of the skin using White Light Interferometry (WLI) on a swatches of skin (1500 μm x 1500 μm) yielded the basic parameters that describe the surface (asperity radii and curvature etc., ).  Figure (9) depicts a typical WLI graph of the skin. The shown inteferogram pertains to a skin spot that is located along the waist of the snake from the belly side (ventral). Two interferograms are depicted: the one to the right hand side of the figure represents the topography of the cell- membrane whereas the one depicted to the left represents a multi-scale scan for the whole skin swatch. Note the scale on the right of the pictures as it indicates the deepest valley and highest point of the skin topography. For these typical skin swatches the value of the deepest part of the membrane was about -120 μm, whereas the highest summit is about 100 μm. The comparable values for the whole swatch are about -5.5 μm and 8.2 μm respectively.

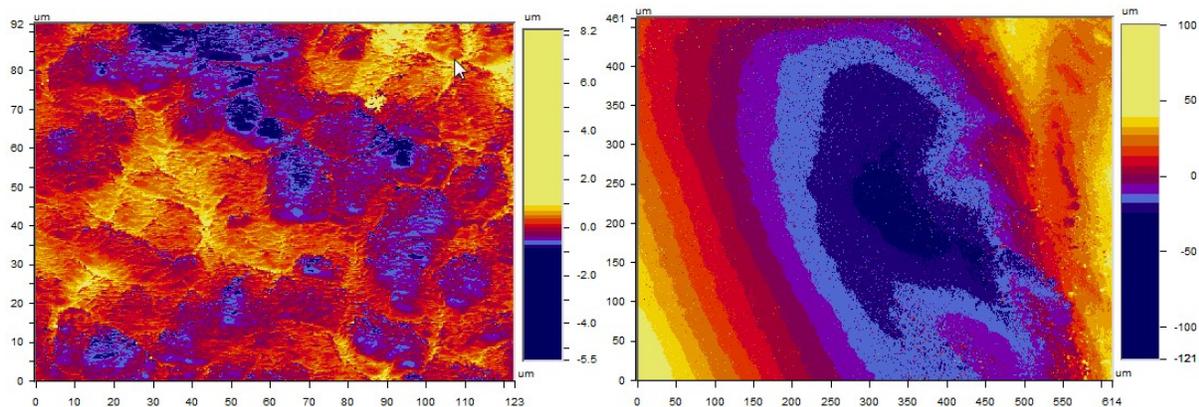

Figure 9   Multi scale WLI graphs depicting the topography of the skin building block (Scale) boundary and membrane (scale bar is in μm).

To establish a measure of comparison between the topography of the snake skin, we compared scans of snake skin to human skin scans. The results are given in figure (10).
Human scans from two different positions are depicted in the figure.  The first scan is from skin located at the back of the hand whereas the second pertains to skin located at the inside of the upper arm of a 40 years old Caucasian female. Samples were obtained by replicating the skin using a replicating silicone. For the snake we chose to compare scans of the ventral scales located in the waist section of the body since it is a major load bearing area during locomotion. Again, it is noted that the peak summit and valley values of the surface of the snake skin are in the order of magnitude of one third that of human skin. Such a comparison highlights, qualitatively, the origins of the superior tribological performance exhibited by the snake.

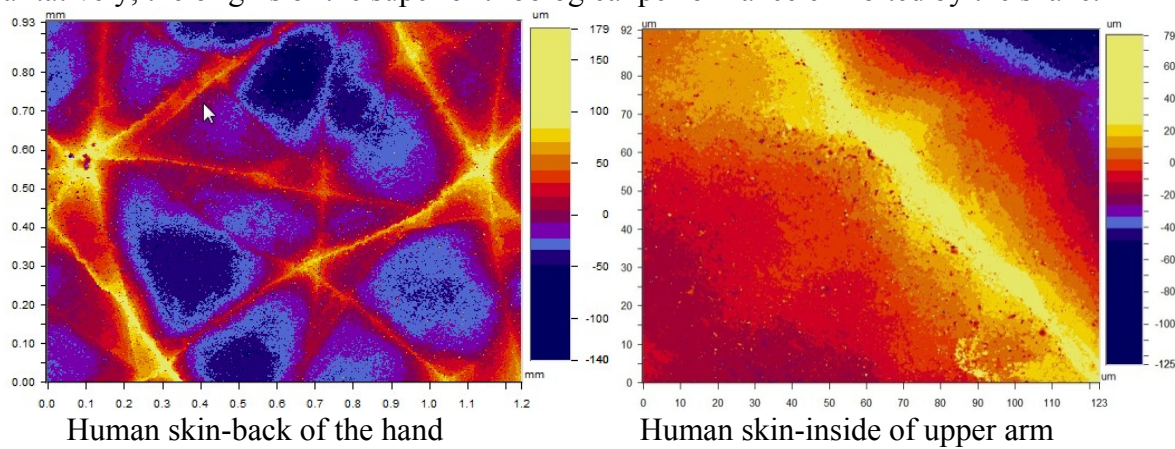

Human skin-back of the hand            Human skin-inside of upper arm

Figure 10  Comparison between the surface texture of snake skin and that of human skin replicas as revealed by White light interferometer (scale bar in µm)

### 3.4. Bearing curve analysis

To complete the analysis, we studied the load bearing characteristics of the skin at each of the predetermined positions (I through IV). Surface parameters were extracted from SEM topography photographs. The complete set of analyzed pictures provided a matrix of roughness parameters that describe the texture of the shed skin at variable scales ranging from X-100 to X-5000. Table 1 (a and b) provides a summary of the parameters extracted from the analysis. It can be seen that the scale of the analysis affects the value of the parameters, which may point at a fractal nature of the surface. Discussion of the implication of such finding is considered out of the scope of this work. However, of interest is to point out one of the features that directly relates to the design of the surface. Comparing the ratios between the Reduced Peak Height *Rpk,* Core Roughness Depth *Rk,* and Reduced Valley Depth *Rvk* reveals symmetry between the positions (compare the columns *Rpk/Rk*, *Rvk/Rk*, and *Rvk/Rpk* of table 1-b, and figure 11). This symmetry is interesting on the count that positions II and III represent the boundaries of the main load bearing regions (trunk). That is the regions on the body where the snake has most of his body weight concentrated (refer to figure 4) and thereby it is the region that is principally used in locomotion. Such symmetry may very well be related to the wear resistance ability of the surface or to the boundary lubrication quality of locomotion. Such a point is a subject of ongoing investigation.

**Table-1 effect of magnification on surface parameters**

a- Surface parameters based on X-250 pictures

|              | Cr/Cf | Cl/Cf | Rpk/Rk | Rvk/Rk | Rvk/Rpk |
|--------------|-------|-------|--------|--------|---------|
| Position I   | 0.718 | 0.861 | 0.391  | 0.159  | 1.144   |
| Position II  | 2.011 | 2.010 | 0.612  | 0.545  | 0.656   |
| Position III | 2.066 | 1.628 | 0.733  | 0.436  | 0.621   |
| Position VI  | 1.388 | 0.926 | 0.617  | 0.195  | 0.749   |

b- Surface parameters based on X-5000 pictures

|              | Cr/Cf | Cl/Cf | Rpk/Rk | Rvk/Rk | Rvk/Rpk |
|--------------|-------|-------|--------|--------|---------|
| Position I   | 1.930 | 1.207 | 0.654  | 0.285  | 0.679   |
| Position II  | 1.273 | 1.803 | 0.478  | 0.404  | 0.812   |
| Position III | 1.671 | 1.622 | 0.484  | 0.359  | 0.800   |
| Position VI  | 1.772 | 1.158 | 0.636  | 0.260  | 0.688   |

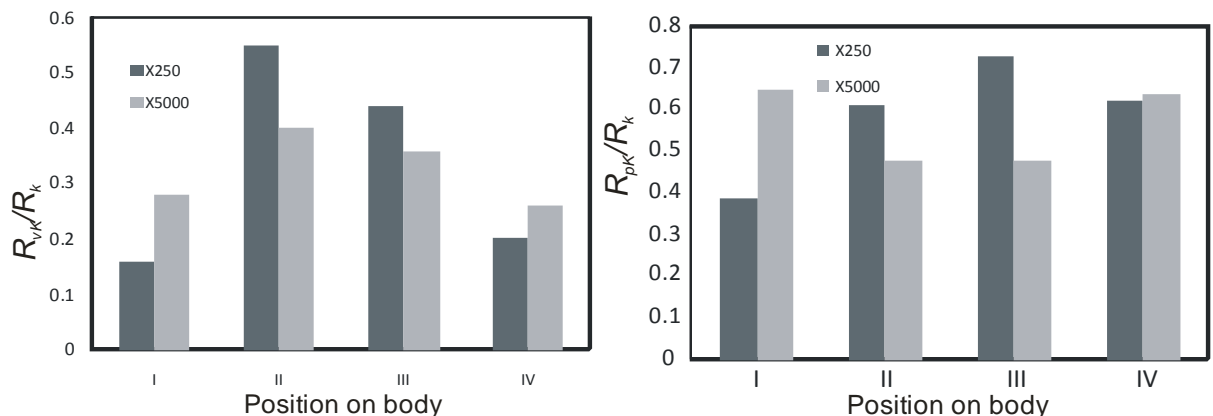

Figure 11: Plot of the ratio of the load bearing parameters Rvk/Rk and Rpk/Rk at two magnifications X-250 and X-5000.

### 4. Friction Behavior

Friction develops between sliding surfaces regardless of the magnitude of the relative motion between them. It acts to resist the relative motion. In doing so, friction transmits, and dissipates energy, to accommodate the various differences between the complying surfaces. The physical processes that contribute to friction have a wide spectrum of length and time scales. Part of the energy dissipation entails converting kinetic energy to thermal energy. A contribution to dissipation also entails acoustical phenomena that involve oscillations and vibration of sliding surfaces.

Friction noises are generally classified in two categories **[11]**. *Mechanical instabilities* such as squeaking of doors, sound radiated by a glass when rubbed by a moist finger, etc,. These stick-slip instabilities are generated under strong contact pressure (i.e., heavily loaded contacts). The second type of friction noise meanwhile, is the so called *roughness noise*. This is generated through dynamical contact of two rough surfaces. The small impacts of opposing asperities generate vibration into the solids. These, in turn, are responsible for the radiated sound. Such a phenomenon occurs under light contact pressure and is characterized by a wideband noise (almost white noise). A pertinent example of such a type of friction noise generation is that of rubbing human fingers on an object (or friction of skin in general). To this end the question of comparing both the friction coefficient of snake skin to that of human skin was raised in the course of the current study. In answering this question one has to be cognizant of the physical and neurophysiological factors that generally affect the friction of skin.

Friction of skin, regardless of the species entails a considerable acoustic contribution which inherently depends on the surface roughness and its evolution during sliding. The physics of roughness noise is related to the dynamics of the so-called multi contact interfaces for which the physical processes involved are not yet fully understood. Several experimental studies **[12-14]** aim to quantify the physical laws governing the process of sound generation the dependence of roughness noise on sliding speed, surface topology and normal load etc,. Although inconclusive, the overall picture emerging from such studies is that all influence factors are of comparable importance with none being strictly dominant.

From a neurophysiological perspective, human skin has a tactile function that may differ from that of reptilian skin. The human skin is provided by various sensors that transform the physical requests applied to surface into electrical impulses. Each of these sensors is sensitive to a particular frequency band. For example, the receptors of Paccini **[15, 16]** are capable of detecting amplitudes of a few microns at a frequency of 300 Hz (maximum of sensitivity). The corpuscles of Meissner, on the other hand, would answer at a lower frequency (50 Hz). Receptors with slow adaptation (Merkell, Ruffini) determine, through a feedback loop system with the brain, the depth of depression of the skin, upon contact, which will inform about the surface tissue stiffness and will determine the slip conditions. In all acoustical emission and friction is a function of the morphological features of the skin and any embedded sensory. Such morphology is not as clear in reptilian skin.

The physical state of the skin also affects friction and sound emission. For example, the density of the keratinocytes of the stratum corneum and the location of the skin within the body is influential. In reptilian skin, location within the body, ventral or dorsal, and the color of the skin (dark or light as in the current species) do affect friction performance and indeed the friction coefficient. Another important factor is the water content of the tested skin. Skin friction is known to be sensitive to moisture content. In all, a thorough answer to the question of comparing tribological performance of both human and python skin, a dedicated detailed study is needed. Never the less for the sake of completeness of our study we have elected to include some of the preliminary COF measurements obtained in our laboratory. It being emphasized that all conclusions drawn, in that respect, will be of qualitative nature.

All measurements utilized a patented tribometer that includes a tribo-acoustic probe sensitive to the range of friction forces and the acoustic emission generated during skin friction. The device is capable of measuring normal, tangential loads and also of detecting sound emission due to sliding. Detailed description of the tribometer is given elsewhere **[17]**. The skin used in measurements consisted of 150 mm long patches taken from four locations on the shed hyde. Skin samples were used as is without any chemical or physical treatment. For each skin patch, measurements were taken for the dorsal and ventral sides of the skin and for patches within the skin of different colors. Four directions of sliding were used: straight forward (SF), straight backward (SB), diagonal forward (DF), and Diagonal backward (DB). Figure (12) depicts the directions used to obtain the measurements and their orientation with respect to the head and tail of the species.

Figure (13) presents a summary plot of the measured COF. Each value is an average of five consecutive measurements taken at the same direction of sliding and identical loading conditions. The results imply that resistance to forward motion is, in general, less than that for moving backwards, observe the difference between the value obtained for the SF direction and that for the SB direction. Anisotropy of the COF is also noted from the results. The COf when moving diagonally is different from that in straight motion. Resistance to motion, however, from tail to head (i.e., forward) is less than that from head to tail (i.e., backward) regardless of the direction of motion (forward or diagonal). These general trends are in line with measurements obtained by other researchers **[18, 19]** who confirmed the anisotropy in friction as well as the increase in resistance to backward motion (although on different species).

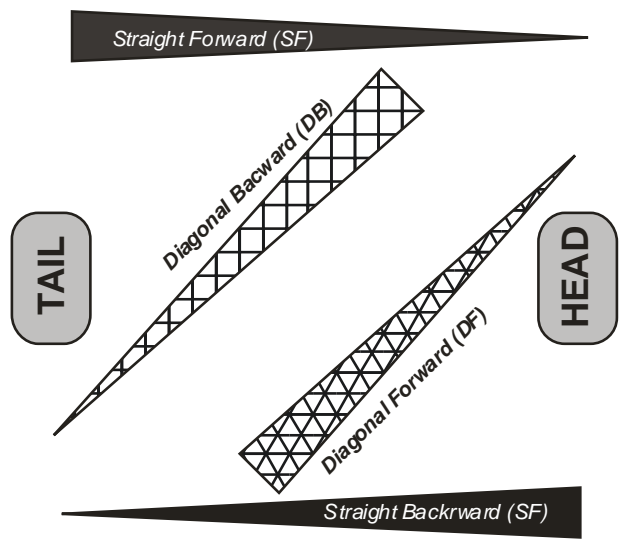

Figure 12  Schematic illustration of the directions of sliding for which coefficient of friction measurements were recorded

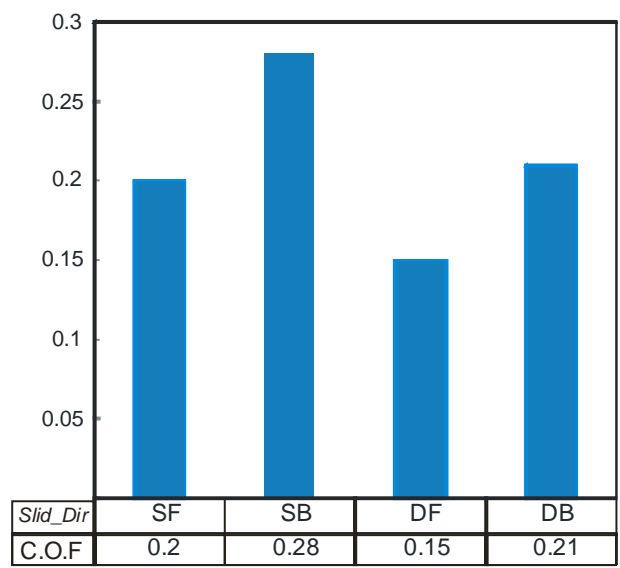

**Figure 13:**  Coefficient of friction values for each of the sliding d irections illustrated in figure 11.

A comparison between the COF of the shed skin to that of human skin extracted from literature **[17]**, is presented in figure (14).  Measurements at two positions are given for human skin, COF of the calf skin and that of the check skin.  Measurements from sliding in the SF and the SB directions for the python skin are compared to those of the human skin.  It can be seen that the COF of snake skin is less than that of human skin.  The difference between the COF of the two skin types is probably more pronounced in actual situations (with the python skin COF less than that for human skin). This because of the difference in moisture content of the skin types used in measurements.  Note that while human COF values were obtained on life human subjects, those obtained for the python employed shed skin which is basically keratinized.

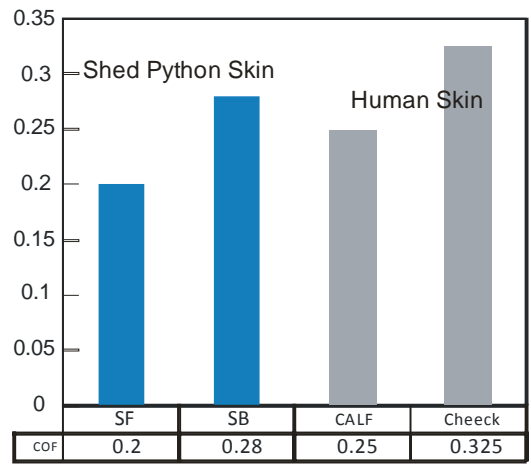

Figure 14  Comparison of COF of human skin, from two zones (calf and cheeck), with the COF of python skin (in SF and SB sliding directions).

5.  **Conclusions and Future Outlook**

In this work we presented the results of an initial study to probe the geometric features of the skin of the Python Regius. It was found the structure of the unit cells is of regionally similar shape (octagonal and hexagonal).

Although almost identical in size and density, the skin constituents (pore density and essential size of the unit cell) vary by position on the body. Comparison of the topography of the snake skin to that of a human female revealed that the surface roughness of the snake species is around one third that of the human samples.

Analysis of the surface roughness parameters implied a multi-scale dependency.  This may point at a fractal nature of the surface a proposition that needs future verification.

The analysis of bearing curve characteristics revealed symmetry between the front and back sections of the body.  It also revealed that the trunk region is bounded by two cross-sections of identical bearing curve ratios.  This has implications in design of textured surfaces that retain an unbreakable boundary lubrication quality and high wear resistance.

The COF of python skin exhibited anisotropic behavior.  The COF in the forward direction was found to be less than that in the backward direction.   Further it was found that human skin has a higher COF than that of Python skin.

Clearly much work is needed to further probe the essential features of the surface geometry. Namely, the basic parametric make up of the topography, form, and their relation to friction and wear resistance.